\documentclass[11pt]{article}
\usepackage{amssymb}
\def\G{\Gamma}
\def\slshD{D \! \! \! \! \slash}
\def\Red{}
\def\Black{}
\def\Blue{}
\begin{document}
\begin{flushright}
{\bf IFUM 680/FT} \\
\end{flushright}
\vskip 1.4 truecm
{\Red
\Large
\bf
\centerline{Refined chiral Slavnov-Taylor identities:}
\centerline{Renormalization and Local Physics}
\normalsize
\rm
\vspace{0.6cm} 
\Black}

\begin{center}
{\Large Marco Picariello \footnote{Marco.Picariello@mi.infn.it} and 
 Andrea Quadri \footnote{Andrea.Quadri@mi.infn.it}}
\\
\vskip 0.5 truecm 
Dipartimento di Fisica, Universit\`a di Milano
via Celoria 16, 20133 Milano, Italy \\
and  INFN, Sezione di Milano \\

\end{center}

\setlength {\baselineskip}{0.2in}
\vskip 0.6 cm
\normalsize

\vskip 1 cm
{\Blue
\centerline{\bf Abstract}
\begin{quotation}
\small
We study the quantization of chiral QED with one family of 
massless fermions and the Stueckelberg field in order to give mass to the
Abelian gauge field in a BRST invariant way. We show
that an extended Slavnov-Taylor (ST) identity can be introduced
and fulfilled to all orders in perturbation theory by a suitable choice
of the local action-like counterterms, order by order in the loop-wise
expansion. This ST identity incorporates the Adler-Bardeen anomaly and involves the introduction 
of a doublet $(K,c)$ where $K$ is an external source of dimension zero and $c$ is the ghost field. 
By a purely algebraic argument
we show that the introduction of the source $K$ trivializes
the cohomology of the extended linearized classical ST operator ${\cal S}'_0$
in the Fadeev-Popov (FP) charge +1 sector.

We discuss the physical content of the extended ST identity and
prove that the cohomology classes associated with ${\cal S}'_0$ are modified
with respect to the ones of the classical BRST differential $s$ in the
FP neutral sector (physical observables). 
This provides a counterexample showing that the introduction of
a doublet can modify the cohomology of the model, as a consequence
of the fact that the counting operator for the doublet $(K,c)$ does not
commute with ${\cal S}'_0$.

We explicitly check that the physical states defined by $s$ are no more
physical states of the full quantized theory
by showing that the subspace of the physical states
corresponding to $s$ is not left invariant under the application of the S matrix,
as a consequence of the extended ST identity.
\end{quotation}
\Black}

\vfill\eject
\section{Introduction}

In perturbative quantum field theory, the full physics can be derived from
the quantum effective action $\G[\phi,\chi]$,
depending on the quantized fields $\phi$ 
and the external sources $\chi$ coupled to local composite operators
${\cal O}(x)$.
$\G[\phi,\chi]$  admits a formal power series expansion in the loop 
parameter~$\hbar$:
\begin{eqnarray}
\G[\phi,\chi] = \sum_{n=0}^{\infty} \hbar^n \G^{(n)}[\phi,\chi] \, .
\label{ij1}
\end{eqnarray}
The zeroth order coefficient $\G^{(0)}$ is identified with the
classical action and it is assumed to be a local functional of
$\phi,\chi$. 

If $\G[\phi,\chi]$ is known, S matrix elements can be computed
by using the LSZ reduction formulae \cite{itzk}. Connected amplitudes are
generated by the Legendre transform $W[J,\chi]$ of $\G[\phi,\chi]$ with respect
to the quantized fields $\phi$:
\begin{eqnarray}
W[J,\chi] = \G[\phi,\chi] + \int d^4x \, \phi J \, , ~~~~
J = - {\delta \G[\phi,\chi] \over \delta \phi} \, .
\label{ij1bis}
\end{eqnarray}
The physics is then recovered by computing the functional derivatives of $W$ with respect to those external sources coupled to physical
composite operators 
at $J=\chi=0$.

If $\G^{(0)}$ is power-counting renormalizable,
the renormalization procedure \cite{velo}
provides a way to compute all 
higher-order terms in the expansion in eq.(\ref{ij1}),
by fixing order by order only a finite set of local action-like 
counterterms.
This procedure is a recursive one, since  it 
allows to construct $\G^{(n)}$ once that 
$\G^{(j)}$, $j <n$ are known. 
From a combinatorial point of view, it turns out that $\G$ is the generating
functional of the 1-PI renormalized Feynman amplitudes.

The behavior of the 
renormalized quantum effective action under an infinitesimal
variation of the quantized fields is 
embodied in the so-called Quantum Action Principle
\cite{QAP}.
It states  that for every local bilinear operator ${\cal S}$, depending
on a set of external sources 
$\phi^*_i$ coupled to local 
operators $\delta \phi_i$ polynomial in the fields and their
derivatives,
the following relation holds true, to all orders in perturbation theory:
\begin{eqnarray}
{\cal S}(\G) \equiv \int d^4x \, \sum_i 
{\delta \G \over \delta \phi^*_i(x)} {\delta \G \over \delta \phi_i(x)} = 
\int d^4x \, \sum_k \left . {\delta \G  \over \delta \zeta_k(x)} 
\right |_{\zeta_k=0} \, .
\label{ij2}
\end{eqnarray}
The operators $\delta \phi_i$ can be identified at the classical level
with the infinitesimal transformations of the 
fields $\phi_i$.
In the R.H.S. of eq.(\ref{ij2}) $\zeta_k$ are external sources
coupled to suitable local composite operators ${\cal O}_k(x)$,
with bounded dimension.
Notice  that in general the external sources can have negative 
dimensions.
Eq. (\ref{ij2}) states that the application of the operator ${\cal S}$
to $\G$ is equivalent to the insertion of the
 set of local
operators ${\cal O}_k(x)$, to all orders in perturbation theory.

In some cases it happens that
\begin{eqnarray}
{\cal S}(\G^{(0)})=0 
\label{ij3}
\end{eqnarray}
but no choice of the local action-like counterterms 
to be fixed order by order in perturbation theory can be made in such a way
that the R.H.S. of eq.(\ref{ij2}) is zero. In this case a classical
symmetry is violated at the quantum level by the R.H.S. of
eq.(\ref{ij2}) and one usually speaks of an anomaly to describe
this kind of behavior of the quantum effective action \cite{PS}.

In actual calculations, the possible breaking terms are considered order by order
in the loop-wise expansion.
Assume then that the effective action has been constructed up to order
$n-1$ so that
\begin{eqnarray}
{\cal S}(\G)^{(j)}=0 \, , ~~~~ j=0,\dots,n-1 .
\label{ij6}
\end{eqnarray}
From eq.(\ref{ij2}) we see that
\begin{eqnarray}
\Delta^{(n)} \equiv {\cal S}(\G)^{(n)}
\label{ij7}
\end{eqnarray}
is a local integrated polynomial in the fields and external sources
and their derivatives, with bounded dimension.

Let us now specialize to gauge theories. We identify
${\cal S}$ with the Slavnov-Taylor operator corresponding to 
the classical BRST symmetry \cite{BRST}.
It turns out in this case that $\Delta^{(n)}$ is further constrained by
a set of consistency conditions \cite{wz}, stemming from
the nilpotency of the BRST transformation.
These consistency conditions are written in a functional way
as
\begin{eqnarray}
{\cal S}_0 \Delta^{(n)}=0 \, ,
\label{ij8}
\end{eqnarray}
where ${\cal S}_0$ denotes the classical linearized
ST operator
\begin{eqnarray}
{\cal S}_0 \Delta^{(n)}
= \int d^4x \, \sum_i \left (
{\delta \G^{(0)} \over \delta \phi^*_i(x)} 
{\delta \Delta^{(n)} \over \delta \phi_i(x)}
+ 
{\delta \Delta^{(n)} \over \delta \phi^*_i(x)} 
{\delta \Gamma^{(0)} \over \delta \phi_i(x)} 
\right ) \, .
\label{ij9}
\end{eqnarray}

Due to the nilpotency of ${\cal S}_0$
and  the locality of $\Delta^{(n)}$, eq.(\ref{ij8})
provides a way to characterize $\Delta^{(n)}$ by studying
the cohomology of ${\cal S}_0$ in the space of 
Lorentz-invariant local
functionals with bounded dimension and Faddeev-Popov (FP) charge $+1$ 
\cite{PS}.
If the only solutions to eq.(\ref{ij8}) are cohomologically trivial,
then it can be shown that the ST identities can be restored at the $n$-th order
by a suitable choice of $n$-th order action-like counterterms.
Otherwise the breaking term $\Delta^{(n)}$ can never be reabsorbed
by a choice of the $n$-th order action-like counterterms, and the theory
is truly anomalous.

We notice that the recursive assumption in eq.(\ref{ij6}) is essential in this process:
if one fails to restore the ST identities at lower orders,
then $\Delta^{(n)}$ actually turns out to be a non-local
functional of the fields and external sources and their
derivatives, hence it cannot be removed by a suitable
choice of $n$-th order action-like counterterms even
for cohomologically non-anomalous theories~\cite{pq}.

In this standard approach of studying which symmetries are
preserved upon quantization, only a small consequence
of the QAP is used, i.e. the locality of the operators
${\cal O}_k(x)$ in the R.H.S. of eq.(\ref{ij2}) is invoked
 to guarantee that $\Delta^{(n)}$
in eq.(\ref{ij7}) is a classical local  polynomial in the fields and external sources and their
derivatives of bounded dimension, provided that eq.(\ref{ij6}) is satisfied.
The full power of the QAP in eq.(\ref{ij2}), i.e. the
fact that the application of the operator ${\cal S}$ to $\G$
is equivalent to the insertion of the set of local operators
 in the R.H.S. of eq.(\ref{ij2}),
to all orders in perturbation theory  and independently
of the action-like local counterterms chosen order by order
in perturbation theory, remains somewhat unexploited.

A more effective, alternative approach to use eq.(\ref{ij2})
would be to regard the anomaly
as a quantum modification of the operator ${\cal S}$,
whose deformation is given by
\begin{eqnarray}
\Delta {\cal S}(\G) \equiv - 
\int d^4x \, \sum_k \left . {\delta \G  \over \delta \zeta_k(x)} 
\right |_{\zeta_k=0} \, .
\label{ij4}
\end{eqnarray}
In this way there is a symmetry obeyed by the quantum effective action
$\G$: 
\begin{eqnarray}
({\cal S} + \Delta {\cal S})(\G)=0 \, .
\label{ij5}
\end{eqnarray}

Recently it has also been pointed out that the introduction of suitably 
defined external sources $\zeta_k$ allows to construct an extended 
linearized classical ST operator, trivializing the cohomology of the model
\cite{barnich}.

In the present paper we apply this approach to chiral QED with one family of 
massless fermions.
We introduce the Stueckelberg field in order to give mass to the
Abelian gauge field in a BRST invariant way.
We study the quantization of the model and 
show that an extended ST identity can be introduced
and fulfilled to all orders in perturbation theory by a suitable choice
of the local action-like counterterms, order by order in the loop-wise
expansion.
This ST identity incorporates the Adler-Bardeen anomaly.

We point out that the corresponding linearized classical ST operator
is nilpotent. However, the physical observables (defined as the cohomology classes
of the linearized classical ST operator in the space of local FP neutral functionals)
turn out to be modified with respect to the ones induced by the classical BRST differential $s$.
Moreover, the space of asymptotic states which are physical
according to the classical BRST differential $s$ 
is not invariant under the S matrix.
This is a consequence of the extended ST identities obeyed by the
quantum effective action $\G$.

\section{Extended ST identities for chiral QED} \label{sec:QED}

We consider the classical Lagrangian of chiral QED with
one family of massless fermions and a massive gauge field $A_\mu$:
\begin{eqnarray}
{\cal L}=  
-{1 \over 4} F_{\mu\nu}^2 +
i \bar \psi \slshD \psi 
+ {1 \over 2} m^2 A_\mu^2 
+ {\alpha \over 2} b^2 - \alpha b \partial A 
+ {1 \over 2} \partial_\mu B \partial^\mu B
- {m^2 \over 2 \alpha} B^2  + \alpha \bar c \square c 
+ m^2 \bar c c
\, .
\label{sez2.1}
\end{eqnarray}
$D_\mu$ is the covariant derivative
\begin{eqnarray}
D_\mu = \partial_\mu - i e {(1 + \gamma^5) \over 2} A_\mu \, .
\label{sez2.2}
\end{eqnarray}
$B$ is the Stueckelberg field \cite{stueckelberg},
 $b$ is the Nakanishi-Lautrup field
\cite{nakanishi}.
$c, \bar c$ are the ghost and antighost fields.
$\alpha$ is the gauge-fixing parameter.

We assign the FP charge by requiring that $A_\mu, \psi, \bar \psi,b,B$
have FP charge $0$, $\bar c$ has FP charge $-1$ and $c$
has FP charge $+1$.

${\cal L}$  is invariant under the following BRST transformations:
\begin{eqnarray}
&& s A_\mu = \partial_\mu c \, , ~~~~
s \bar c = b + {m \over \alpha} B \, , ~~~~ sb = -{m^2 \over \alpha} c \, ,  
~~~~ s B= mc \, , \nonumber \\
&& s \psi = -i e  {(1+\gamma^5) \over 2} \psi c \, , ~~~~
s \bar \psi = - i e  c \bar \psi {(1-\gamma^5) \over 2} \, , ~~~~
s c =0 \, .
\label{sez2.3}
\end{eqnarray}
$s$ is nilpotent.
In order to define at the quantum level the composite operators
$s \psi, s \bar \psi$ appearing in eq.(\ref{sez2.3}) one has to
couple them in the classical action 
$\G^{(0)}$ to classical external sources  
$\bar \eta, \eta$ 
(known as antifields in the Batalin-Vilkovisky
formalism \cite{GPS}):
\begin{eqnarray}
\G^{(0)} = \int d^4x \, \left ( {\cal L} -ie 
\bar \eta {(1+\gamma^5) \over 2} \psi c 
 - i e  c \bar \psi {(1-\gamma^5) \over 2} \eta \right ) \, .
\label{azclass}
\end{eqnarray}
The invariance of 
$\G^{(0)}$ under
the BRST differential $s$ is now expressed as \cite{zj}
\begin{eqnarray}
&& {\cal S}(\G^{(0)}) \equiv 
\int d^4x \, 
\left (
\partial_\mu c {\delta \G^{(0)} \over \delta A_\mu}
+ \left ( b + {m \over \alpha}B \right ) 
 {\delta \G^{(0)} \over \delta \bar c}
+ mc {\delta \G^{(0)} \over \delta B}
- {m^2 \over \alpha}c {\delta \G^{(0)} \over \delta b} 
\right . \nonumber \\
&& \left . 
~~~~~~~~~~~~~~~~~~~~~~~~ + {\delta \G^{(0)} \over \delta \bar \eta}
{\delta \G^{(0)} \over \delta \psi}
+ {\delta \G^{(0)} \over \delta \eta} 
{\delta \G^{(0)} \over \delta \bar \psi}
\right ) =  0 \, .
\label{sez2.6}
\end{eqnarray}
In the above equation we have introduced the ST operator ${\cal S}$.
The bilinear part is only reduced to the fermion sector since
the BRST variations of $A_\mu, \bar c$, being linear in the quantized
fields, do not require the introduction of the corresponding antifields.
The requirement that $\G^{(0)}$ is FP neutral implies that 
the antifields $\eta ,\bar \eta$ carry FP charge $-1$.

The dependence of $\G^{(0)}$ on $b,B,\bar c$ is given by 
\begin{eqnarray}
{\delta \G^{(0)} \over \delta b} = \alpha b - \alpha \partial A \, , ~~~~~
{\delta \G^{(0)} \over \delta B} = - \left ( \square  + { m^2 \over \alpha} 
\right ) B 
\label{new1}
\end{eqnarray}
and
\begin{eqnarray}
{\delta \G^{(0)} \over \delta \bar c} = \alpha \square c + m^2 c \, .
\label{new3}
\end{eqnarray}
The last equation is known as the classical ghost equation. 
The R.H.S. of eqs.(\ref{new1}-\ref{new3}) is linear in the quantized fields, hence
we can preserve the above equations at the quantum level by a suitable choice of local action-like counterterms, order by order in the perturbative expansion.

So we require that the full quantized effective action $\G$ satisfies
the conditions:
\begin{eqnarray}
{\delta \G \over \delta b} = \alpha b - \alpha \partial A  \, , ~~~~~ 
{\delta \G \over \delta B} = - \left ( \square  + { m^2 \over \alpha} 
\right ) B 
\label{new1.bis}
\end{eqnarray}
and
\begin{eqnarray}
{\delta \G \over \delta \bar c} = \alpha \square c + m^2 c\, .
\label{new3.bis}
\end{eqnarray}
The second of eqs.(\ref{new1.bis}) entails that $B$ is a free field.
By eq.(\ref{new3.bis}) the ghost field decouples.

By the QAP and eq.(\ref{sez2.6}), the first order ST breaking term
\begin{eqnarray}
\Delta^{(1)} \equiv {\cal S}(\G)^{(1)}
\label{new4}
\end{eqnarray}
is a Lorentz-invariant integrated polynomial
in the fields and the external sources and their derivatives
with dimension less or equal to $5$ and FP charge $1$. 
$\Delta^{(1)}$ satisfies the Wess-Zumino consistency condition
\begin{eqnarray}
{\cal S}_0( \Delta^{(1)} ) = 0 \, 
\label{new5}
\end{eqnarray}
where ${\cal S}_0$ is the classical linearized ST operator given by
\begin{eqnarray}
{\cal S}_0 & = &
\int d^4x \, 
\left (
\partial_\mu c {\delta \over \delta A_\mu}
+ \left (b + {m \over \alpha} B \right )  {\delta  \over \delta \bar c}
+ mc {\delta \over \delta B} - {m^2 \over \alpha} c {\delta \over \delta b}
\right . 
\nonumber \\
&& ~~~~~~~~~~~
\left . 
+ {\delta \G^{(0)} \over \delta \bar \eta}
{\delta \over \delta \psi}
+  
{\delta \G^{(0)} \over \delta \psi} {\delta \over \delta \bar \eta} 
+ {\delta \G^{(0)} \over \delta \eta} 
{\delta \over \delta \bar \psi}
+ 
{\delta \G^{(0)} \over \delta \bar \psi} {\delta \over \delta \eta} 
\right )  \, .
\label{z4bis}
\end{eqnarray}
The solution of eq.(\ref{new5}) is \cite{anomalies}
\begin{eqnarray}
\Delta^{(1)} = r
 \int d^4x  \, c \, 
\epsilon_{\mu \nu \rho \sigma} \partial^\mu A^\nu 
\partial^\rho A^\sigma + {\cal S}_0 (\Xi^{(1)}) 
\label{new6}
\end{eqnarray}
for some local action-like functional  $\Xi^{(1)}$. 
The breaking term ${\cal S}_0 (\Xi^{(1)})$ can be reabsorbed by adding
to $\G^{(1)}$ the counterterm functional $-\Xi^{(1)}$. This amounts
to change the first order normalization conditions \cite{FGQ}.

$\Delta^{(1)}$ is thus reduced to the following
Adler-Bardeen chiral anomaly:
\begin{eqnarray}
\Delta^{(1)}_{\rm AB} = r
 \int d^4x  \, c \,
\epsilon_{\mu \nu \rho \sigma} \partial^\mu A^\nu 
\partial^\rho A^\sigma \,.
\label{z2}
\end{eqnarray}
for some non-zero c-number $r$.
$r$ cannot be set equal to zero
by any choice of the local action-like first order counterterms
in $\G^{(1)}$.
The  ST breaking in eq.(\ref{z2}) is an anomalous one.
As a consequence, the second order ST breaking term $\Delta^{(2)} = {\cal S}(\G)^{(2)}$ turns out to be
a non-local functional of the fields and external sources and their derivatives \cite{pq}.

Due to the Abelian character of the model,
we can write from eq.(\ref{z2})
\begin{eqnarray}
{\cal S}(\G)^{(1)} =  r
 \int d^4x  \, c \,
\epsilon_{\mu \nu \rho \sigma} \partial^\mu A^\nu 
\partial^\rho A^\sigma = 
\int d^4x \, c {\delta \G^{(1)} \over \delta K} \, .
\label{z3}
\end{eqnarray}
This is only possible since the ghost field decouples in the Abelian case,
due to the ghost equation in eq.(\ref{new3.bis}).
$K$ is an external source coupled in $\G^{(1)}$ to the Adler-Bardeen term 
$\epsilon_{\mu \nu \rho \sigma} \partial^\mu A^\nu \partial^\rho A^\sigma$.
Notice that $K$ has dimension $0$.

In the spirit of eq.(\ref{ij5}), we can deform the operator ${\cal S}$ 
into ${\cal S}'$ defined by
\begin{eqnarray}
{\cal S}' \equiv {\cal S} + \int d^4x \, c {\delta \over \delta K} \, .
\label{new9}
\end{eqnarray}
It then follows that
\begin{eqnarray}
{\cal S}' (\G)^{(1)}=0 \, .
\label{new10}
\end{eqnarray}
One should regard the operator ${\cal S}'$
in eq.(\ref{new10}) as defining the
(extended) chiral symmetry of the model.

We now show that the ST identity associated with ${\cal S}'$ can
be restored to all orders in perturbation theory.
The proof is a recursive one.

For $n=0$  we have
\begin{eqnarray}
{\cal S}'(\G^{(0)}) = 0 
\label{new11}
\end{eqnarray}
since $\G^{(0)}$ does not depend on $K$.
For $n=1$ the extended ST identity is satisfied (see eq.(\ref{new10})).
Assume that it is fulfilled till order $n-1$:
\begin{eqnarray}
{\cal S}'(\G)^{(j)}=0 \, , ~~~~ j=0,1,\dots,n-1 \, .
\label{new11.bis}
\end{eqnarray}
We shall prove that
the extended ST identity can be fulfilled at the $n$-th order by
a suitable choice of the $n$-th order local counterterms.
We point out that, since $K$ has dimension zero, power counting arguments
cannot be effectively 
used to constrain the dependence of $\Delta^{(n)}$ on $K$.

We first notice that the following algebraic identity holds true
for every bosonic functional $\G$:
\begin{eqnarray}
{\cal S}'_{\G} ({\cal S}'(\G)) = 0 \, ,
\label{new12}
\end{eqnarray}
where ${\cal S}'_{\G}$ is the linearized extended ST operator for $\G$:
\begin{eqnarray}
&& {\cal S}'_\G = 
\int d^4x \, 
\left (
\partial_\mu c {\delta \over \delta A_\mu}
+ \left ( b + {m \over \alpha} B \right )
 {\delta  \over \delta \bar c}
+mc {\delta \over \delta B}
- {m^2 \over \alpha} c {\delta \over \delta b}
\right .
\nonumber \\
&&
\left . 
~~~~~~~~~~~~~~~~~~
+ {\delta \G \over \delta \bar \eta}
{\delta \over \delta \psi}
+ 
{\delta \G \over \delta \psi}
 {\delta \over \delta \bar \eta}
+ {\delta \G \over \delta \eta} 
{\delta \over \delta \bar \psi}
+ 
{\delta \G \over \delta \bar \psi}
{\delta \over \delta \eta} 
+ c {\delta \over \delta K}
\right )  \, .
\label{new13}
\end{eqnarray}
By using eq.(\ref{new12}) and eq.(\ref{new11.bis}) 
we obtain the following Wess-Zumino
consistency condition for the $n$-th order breaking term
$\Delta^{(n)} \equiv {\cal S}'(\G)^{(n)}$:
\begin{eqnarray}
{\cal S}'_0 (\Delta^{(n)}) = 0 \, ,
\label{new14}
\end{eqnarray}
where ${\cal S}'_0$ is the extended classical linearized ST operator
given by
\begin{eqnarray}
{\cal S}'_0 & = &
\int d^4x \, 
\left (
\partial_\mu c {\delta \over \delta A_\mu}
+ \left ( b + {m \over \alpha} B \right )
{\delta  \over \delta \bar c}
+mc {\delta \over \delta B}
- {m^2 \over \alpha} c {\delta \over \delta b}
\right . \nonumber \\
&& ~~~~~~~~~~  \left . 
+ {\delta \G^{(0)} \over \delta \bar \eta}
{\delta \over \delta \psi}
+ 
{\delta \G^{(0)} \over \delta \psi}
 {\delta \over \delta \bar \eta}
+ {\delta \G^{(0)} \over \delta \eta} 
{\delta \over \delta \bar \psi}
+
{\delta \G^{(0)} \over \delta \bar \psi}
 {\delta \over \delta \eta} 
+ c {\delta \over \delta K}
\right )  \nonumber \\
& = & {\cal S}_0 + \int d^4x \, c {\delta \over \delta K} \, .
\label{new15}
\end{eqnarray}
Notice that $\left . {\cal S}'_0 \right . ^2=0$.
By the QAP, $\Delta^{(n)}$ is a local functional of the fields
and external sources with dimension less or equal to $5$ and
FP charge $+1$.

By explicit computation it can be shown by using eqs.(\ref{new1.bis}) and
(\ref{new3.bis}) that $\Delta^{(n)}$ does not depend on $b,B,\bar c$ and
is a functional of $A^\mu,c,\bar \psi, \psi, \eta, \bar \eta$
only.

Moreover, by power counting $\Delta^{(n)}$ cannot depend on $\eta, \bar \eta$.
Having ruled out the dependence on all fields and external sources
with negative FP charge, we see that $\Delta^{(n)}$ depends on the
ghost field $c$ only linearly:
\begin{eqnarray}
\int d^4x \, c {\delta \Delta^{(n)} \over \delta c} = \Delta^{(n)} \, .
\label{new17}
\end{eqnarray}
We notice that $\Delta^{(n)}$ is a polynomial with respect
to the quantized fields $A^\mu, \psi, \bar \psi$, since they have
positive dimension. However, it can be a truly formal power series
in the dimensionless external source $K$.

We now show that $\Delta^{(n)}$ is 
the ${\cal S}'_0$-image of a local action-like functional
(i.e. the cohomology of ${\cal S}'_0$ is empty).
For this purpose we introduce the counting operator for $K$:
\begin{eqnarray}
{\cal N} = \int d^4x \, K {\delta \over \delta K}
\label{new18}
\end{eqnarray}
and decompose $\Delta^{(n)}$ according to the degree induced by ${\cal N}$:
\begin{eqnarray}
\Delta^{(n)}=\sum_{j=0}^\infty \Delta^{(n)}_j \, , ~~~~ {\cal N} \Delta^{(n)}_j = j \Delta^{(n)}_j \, .
\label{new19}
\end{eqnarray}

We can always reabsorb $\Delta^{(n)}_0$ in eq.(\ref{new19}). Indeed,
by standard cohomological results \cite{PS,anomalies} 
it is known that $\Delta^{(n)}_0$ can always be reduced  by adding suitable local $n$-th order
counterterms independent of $K$ to
the Adler-Bardeen term:
\begin{eqnarray}
\Delta^{(n)}_0 = r^{(n)}  \int d^4x \, c \epsilon_{\mu \nu \rho \sigma} \partial^\mu A^\nu
\partial^\rho A^\sigma \, .
\label{new20.bis}
\end{eqnarray}
This is compensated by the counterterm
\begin{eqnarray}
\Xi^{(n)}_0 \equiv - r^{(n)} \int d^4x \, K  \epsilon_{\mu \nu \rho \sigma} \partial^\mu A^\nu
\partial^\rho A^\sigma \, .
\label{new20.ter}
\end{eqnarray}

Assume now that $\Delta^{(n)}_j =0$ for $j=0,\dots,m-1$. We show that one can add suitably defined
local action-like counterterms to $\G^{(n)}$ in such a way that $\Delta^{(n)}_m$ is also zero.

For this purpose we define the operator~\cite{ZUMINO}
\begin{eqnarray}
{\cal H} = \int_0^1 dt \int d^4x \,
K \lambda_t {\delta \over \delta c} \, .
\label{ch}
\end{eqnarray}
In the previous equation we have introduced the operator $\lambda_t$ given by
\begin{eqnarray}
\lambda_t \Delta^{(n)}(c,K,\varphi) = \Delta^{(n)}(tc, tK,\varphi) \, ,
\label{new19b.bis}
\end{eqnarray}
where we have denoted by $\varphi$ all fields and external sources other than $c,K$ on which
$\Delta^{(n)}$ might depend.
${\cal H}$
is a contracting homotopy for the differential 
\begin{eqnarray}
\sigma = \int d^4x \,  c {\delta \over \delta K} \, ,
\label{diff}
\end{eqnarray}
since
\begin{eqnarray}
\{ {\cal H}, \sigma \} \Delta^{(n)}(K,c,\varphi)  & = &
\int_0^1  dt \int d^4x \, 
\left ( K \lambda_t {\delta \over \delta K} + c \lambda_t {\delta \over \delta c} 
\right ) \Delta^{(n)}(K,c,\varphi) \nonumber \\
& = & 
 \Delta^{(n)}(K,c,\varphi)  - \Delta^{(n)}(0,0,\varphi) \, .
\label{motivo}
\end{eqnarray}
We now notice that the following identity holds true for any 
functional $\Delta^{(n)}$ obeying eq.(\ref{new17}):
\begin{eqnarray}
\{ {\cal H}, \sigma \} \Delta^{(n)}(K,c,\varphi) = 
{\cal S}'_0 \left ( {\cal H} \Delta^{(n)} \right )  - 
{\cal S}_0 \left( {\cal H} \Delta^{(n)}\right)
 \, .
\label{new20_1}
\end{eqnarray}
so that  by eq.(\ref{motivo})
\begin{eqnarray}
\Delta^{(n)}(K,c,\varphi) - \Delta^{(n)}(0,0,\varphi) = 
{\cal S}'_0 \left ( {\cal H} \Delta^{(n)} \right )  - 
{\cal S}_0 \left( {\cal H} \Delta^{(n)}\right)
 \, .
\label{new20}
\end{eqnarray}

By assumption we can write
\begin{eqnarray}
\Delta^{(n)}= \sum_{k \geq m} \Delta^{(n)}_k \, .
\label{new21}
\end{eqnarray}
Inserting eq.(\ref{new21}) in eq.(\ref{new20}) we obtain
\begin{eqnarray}
&& \Delta^{(n)}(K,c,\varphi)
- \Delta^{(n)}(0,0,\varphi) = \sum_{k \geq m} \Delta^{(n)}_k 
= {\cal S}'_0( 
{\cal H} \Delta^{(n)} )
- {\cal S}_0( 
{\cal H} \Delta^{(n)} 
) 
\label{new22}
\end{eqnarray}
and finally
\begin{eqnarray}
\sum_{k \geq m} \Delta^{(n)}_k - {\cal S}'_0\left( 
{\cal H} \Delta^{(n)}
\right)
=  -{\cal S}_0\left(
{\cal H} \Delta^{(n)}
\right) \, .
\label{new22.bis}
\end{eqnarray}
Since ${\cal S}_0$ does not depend on $K$, the R.H.S. of eq.(\ref{new22.bis})
 admits
an expansion according to the degree induced by ${\cal N}$ starting from 
$m+1$.
Hence $\Delta^{(n)}_m$ can be set equal to zero by adding to $\G^{(1)}$ 
the counterterm
\begin{eqnarray}
\Xi^{(n)}_m \equiv {\cal H} \Delta^{(n)} = 
- \int d^4x \int_0^1 dt \, K \lambda_t {\delta \Delta^{(n)} \over \delta c} \, .
\label{new23}
\end{eqnarray}

This concludes the proof that the extended ST identity can be restored to all orders
in perturbation theory.


\section{Physical observables} \label{sez3}

We have shown that the quantum effective action $\G$ satisfies the
extended ST identity
\begin{eqnarray}
{\cal S}'(\G)=0 \, .
\label{s3.1}
\end{eqnarray}
Moreover, the corresponding linearized classical ST operator
${\cal S}'_0$ is nilpotent. 
We now investigate the consequence
on the physics of the extended ST identity in eq.(\ref{s3.1}).

We identify physical observables with the cohomology classes
of the extended linearized ST operator ${\cal S}_0'$ in the
space of local functionals with FP charge $0$.
The physical observables generated by ${\cal S}_0'$ are
different from the ones obtained from ${\cal S}_0$.
In the present model the latter coincide with the ones obtained from $s$
\cite{anomalies}.
We first work out an example and then discuss the general
situation.

We consider the gauge mass term 
\begin{eqnarray}
{\cal M} = \int d^4x \,  {1 \over 2}m^2 A_\mu^2 \, .
\label{dd1}
\end{eqnarray}
${\cal M}$ is neither a ${\cal S}_0$-invariant nor 
a ${\cal S}'_0$-invariant. However, the functional
\begin{eqnarray}
{\cal M}_K \equiv {\cal M} + m^2 \int d^4x \, \left ( K \partial A - {1 \over 2} K \square K \right )
\label{dd2}
\end{eqnarray}
is a ${\cal S}'_0$-invariant. By explicit computation it can be
verified that ${\cal M}_K$ 
is not the ${\cal S}_0'$-variation of any local functional
with FP charge $-1$,
hence it identifies a true observable of the theory.
Notice that ${\cal M}$
 is not a ${\cal S}_0$-invariant, thus it does not belong to the 
cohomology of ${\cal S}_0$ in the space of local functionals with FP charge zero.

We now go on shell by imposing the conditions
\begin{eqnarray}
J = -{\delta \G \over \delta \phi} =0 \, , ~~~~ {\chi} = 0 \, .
\label{d5_1}
\end{eqnarray}
The second of the conditions in eqs.(\ref{d5_1}) entails $K=0$,
hence on-shell the representative ${\cal M}_K$ has to be identified
with ${\cal M}$: the mass term for the gauge field is an on-shell observable
of the model. 

This mechanism of extension of the cohomology due to the introduction of the source $K$ 
applies to the whole FP neutral sector of local functionals.
Let ${\cal G}$ be a local functional with FP charge $0$.
Then $\Delta' \equiv {\cal S}'_0({\cal G})$ is a local functional with FP 
charge $+1$
and by the arguments of the previous section there exists a local functional $R$ depending
on $K$ and all other fields and external sources of the model such that
\begin{eqnarray}
\Delta' = {\cal S}'_0({\cal R}) \, .
\label{agg1}
\end{eqnarray}
${\cal R}$ is not uniquely defined. Notice in particular that
${\cal R}$ can be chosen in such a way that $\left . {\cal R} \right |_{K=0}=0$.
This follows from the arguments of sect. \ref{sec:QED}.

Hence
\begin{eqnarray}
{\cal S}'_0({\cal G} -{\cal R}) = 0 \, .
\label{agg2}
\end{eqnarray}
This means that 
\begin{eqnarray}
{\cal G}_K \equiv {\cal G} - {\cal R}
\label{d4}
\end{eqnarray}
is a representative of the cohomology class of a local 
observable ${\cal O}$.

 Going on-shell ${\cal G}_K$ reduces to 
${\cal G}$, hence in the extended theory governed by ${\cal S}'_0$
every functional whose ${\cal S}'_0$-variation is non-zero
is actually a representative of an on-shell local observable.

We conclude that the physical content of the quantized theory governed 
by ${\cal S}'_0$ has changed with respect to the classical
theory, whose physics is described by the local FP neutral cohomology classes of the
classical BRST differential $s$.

This can also be checked by studying the time-evolution of the asymptotic states
which are physical according to the classical BRST differential  $s$.

We follow the technique discussed in \cite{becchi}. 
According
to the reduction formulae the connected S matrix can be expressed as
\begin{eqnarray}
S = : \left . \Sigma : W[J,\chi] \right |_{J=\chi=0}
\label{s3.2}
\end{eqnarray}
where the operator $\Sigma$ is defined by
\begin{eqnarray}
:\Sigma: = : \exp  
\left ( 
\int d^4x d^4y \, \varphi_i(x) \G_{ij}(x-y) {\delta \over \delta J_j(y)}  
\right ) : \, ,
\label{s3.3}
\end{eqnarray}
and $::$ stands for normal ordering.

$\varphi_i$ are linear combinations of the asymptotic fields 
$\phi^{as}_i$
\begin{eqnarray}
\varphi_i = a_{ij} \phi^{as}_j \, ,
\label{s3.3bis}
\end{eqnarray}
where the matrix $a_{ij}$ is invertible \cite{becchi}.
In eq.(\ref{s3.3}) we have denoted by subscripts 
the functional differentiation with respect to the arguments of
$\G[\phi,\chi]$ evaluated at $\phi=\chi=0$:
\begin{eqnarray}
\G_{ij}(x-y) \equiv
\left . {\delta^2 \G \over \delta \phi_i(x) \phi_j(y)} \right |_{\phi=\chi=0}
\, . 
\label{s3.4}
\end{eqnarray}
In the following
$J_c$ denotes the external source 
coupled to $c$, $J_\mu$ the source coupled to $A_\mu$, $J_{\bar c}$ the one
coupled to $\bar c$ and so on.

The extended ST identity on the connected generating functional $W[J,\chi]$
reads
\begin{eqnarray}
{\cal S}'(W) = {\cal S}(W) - \int d^4x \, {\delta W \over \delta J_c}
{\delta W \over \delta K} = 0 \, ,
\label{s3.5}
\end{eqnarray}
with ${\cal S}(W)$ given by
\begin{eqnarray}
{\cal S}(W) & = & - \int d^4x \, 
\left ( \partial_\mu {\delta W \over \delta J_c} J_\mu 
+ \left ( {\delta W \over \delta J_b} + 
{m \over \alpha} {\delta W \over \delta J_B} \right ) J_{\bar c}
+ m {\delta W \over \delta J_c} J_B \right . \nonumber \\
& & ~~~~~~~~ \left . 
- {m^2 \over \alpha} {\delta W \over \delta J_c} J_b
+ {\delta W \over \delta \bar \eta} J_{\psi} 
+ {\delta W \over \delta \eta} J_{\bar \psi} 
\right ) \, .
\label{s3.6}
\end{eqnarray}

We introduce the operator $Q$ \cite{CF} acting on the fields $\varphi_i$ 
defined by
\begin{eqnarray}
& [Q, A_\mu ] = \partial_\mu c \, , ~~~~
\{Q, c\} =0 \, , ~~~~ \{Q, \bar c\} = b + {m \over \alpha}B \, , ~~~~
[Q, B] = mc \, , & \nonumber \\
& [Q,b] = -{m^2 \over \alpha} c \, , ~~~~ \{ Q, \psi \} = 0 \, ,
 ~~~~ \{ Q,\bar\psi \}= 0 \, . &
\label{s3.9}
\end{eqnarray}
$Q$ expresses the action induced by the classical BRST differential $s$ on the fields $\varphi_i$.
Notice that $Q$ is nilpotent. The physical subspace corresponding to the cohomology generated by 
the classical BRST differential $s$
can be identified with ${\rm ker \ Q} / {\rm Im \ Q}$.
However, $Q$ does not commute with the
S matrix.

For that purpose, 
we first compute $[Q, :\Sigma:]$ and get (in the
momentum space representation)
\begin{eqnarray}
[Q, :\Sigma:]  & = &
:\!\!\!\!\int d^4p\left[ 
c \left ( i p^\mu \G_{\mu\nu} - {m^2 \over \alpha} \G_{b\nu}
+ m \G_{B\nu} \right ) {\delta \over \delta 
J_\nu} + c \left ( i p^\mu \G_{\mu B} - {m^2 \over \alpha} \G_{bB}
+ m \G_{BB} \right ) 
{\delta \over \delta J_B} \right . \nonumber \\
&& ~~~~~~~~~~ \left . 
+ c \left (i p^\mu \G_{\mu b} -{m^2 \over \alpha} \G_{bb} + m \G_{Bb}  \right )
{\delta \over \delta J_b} + \left ( b
+ {m \over \alpha} B \right )  \G_{\bar c c} {\delta \over \delta  J_c} 
\right ] \Sigma: \nonumber \\
& = & 
: \int d^4p \, \left [ 
c \left ( i p^\mu \G_{\mu\nu} - {m^2 \over \alpha} \G_{b\nu} \right ) 
{\delta \over \delta 
J_\nu} + c \, m \G_{BB} 
{\delta \over \delta J_B} \right . \nonumber \\
&& ~~~~~~~~~~ \left . 
+ c \left (i p^\mu \G_{\mu b} -{m^2 \over \alpha} \G_{bb}   \right )
{\delta \over \delta J_b} + \left ( b
+ {m \over \alpha} B \right )  \G_{\bar c c} {\delta \over \delta  J_c} 
\right ] \Sigma: \, .
\label{s3.10}
\end{eqnarray}
In the second line of the above equation we have taken into account
eqs.(\ref{new1.bis}).

We then use the extended ST identity to constrain the two-point
functions appearing in eq. (\ref{s3.10}):
\begin{eqnarray}
&& \left . {\delta ^2 {\cal S}'(\G) \over \delta c \delta A_\nu} 
\right |_{\phi=\chi=0} =
i p^\mu \G_{\mu\nu} - {m^2 \over \alpha} \G_{\nu b} 
+ \G_{\nu K} = 0 \, , \nonumber \\
&&  \left . {\delta ^2 {\cal S}'(\G) \over \delta c \delta B} 
\right |_{\phi=\chi=0} = m \G_{BB} + {m \over \alpha} \G_{c \bar c} = 0 \, ,
\nonumber \\
&&  \left . {\delta ^2 {\cal S}'(\G) \over \delta c \delta b} 
\right |_{\phi=\chi=0} =
i p^\mu \G_{\mu b} + \G_{c \bar c} - {m^2 \over \alpha} \G_{bb} = 0 \, ,
\label{s3.11}
\end{eqnarray}
by using again eqs.(\ref{new1.bis}).
We now insert eqs.(\ref{s3.11}) into eq.(\ref{s3.10})
and get
\begin{eqnarray}
\hspace{-14mm} && [Q, :\Sigma:]  =  
: \int d^4p \, \left [ 
-c \G_{\nu K} {\delta \over \delta J_\nu} + 
b \G_{\bar c c} {\delta \over \delta J_c}
 - c \G_{c \bar c} {\delta \over \delta J_b}
- {m \over \alpha}c \G_{c \bar c} {\delta \over \delta J_B} 
+ {m \over \alpha} B \G_{\bar c c} {\delta \over \delta J_c}
\right ] \Sigma :
\label{s3.12}
\end{eqnarray}
One also gets
\begin{eqnarray}
[: \Sigma :, {\cal S}] = 
: \int d^4p \, \left [ 
\left ( -A^\mu \G_{\mu K}  + 
b \G_{\bar c c} \right ) {\delta \over \delta J_c}
 - c \G_{c \bar c} {\delta \over \delta J_b}
-{m \over \alpha} B \G_{\bar c c} {\delta \over \delta J_c}
-{m \over \alpha} c \G_{c \bar c} {\delta \over \delta J_B}
\right ] \Sigma :
\label{s3.13}
\end{eqnarray}
This follows by explicit computation once that eqs.(\ref{s3.11}) 
are taken into account.
By combining eqs.(\ref{s3.12}) and (\ref{s3.13}) we 
get
\begin{eqnarray}
[Q , : \Sigma: ] & = & [: \Sigma :, {\cal S} ] 
+ : \int d^4p \left ( A^\mu \G_{\mu K} {\delta \over \delta J_c} 
-c \G_{\nu K} {\delta \over \delta J_\nu}
\right ) \Sigma : \, . 
\label{s3.14}
\end{eqnarray}
We now notice that
\begin{eqnarray}
\left . \left [ : \Sigma :, {\cal S} \right ] (W) \right |_{J=\chi=0} & = &
\left . : \Sigma : \left ( \int d^4x {\delta W \over \delta J_c} {\delta W \over \delta K} \right )
\right |_{J=\chi=0} \, .
\label{s3.7}
\end{eqnarray}
In the Abelian case we can use the ghost equation in eq.(\ref{new3.bis})
 to obtain 
from eq.(\ref{s3.7})
\begin{eqnarray}
\left . \left [ : \Sigma :, {\cal S} \right ] (W) \right |_{J=\chi=0} & = &
\left . : \int d^4x \, c(x) {\delta \over \delta K(x)} \Sigma : W 
\right |_{J=\chi=0} \, .
\label{s3.8}
\end{eqnarray}
Hence by eq.(\ref{s3.8})
\begin{eqnarray}
[Q, S] = \left . [Q , : \Sigma: ] W \right |_{J=\chi=0} =
\left .  : \int d^4p \left ( A^\mu \G_{\mu K} {\delta \over \delta J_c} 
-c \G_{\nu K} {\delta \over \delta J_\nu} + c {\delta \over \delta K}
\right ) \Sigma : W\right |_{J=\chi=0} \, .
\label{s3.15}
\end{eqnarray}
The R.H.S. of eq.(\ref{s3.15}) is zero if 
\begin{eqnarray}
{\delta W \over \delta K} = {\delta \G \over \delta K} = 0 \, .
\label{s3.16}
\end{eqnarray}
The above equation is satisfied at tree level, but cannot hold
true at the quantum level due to the appearance of the Adler-Bardeen
anomaly. 
We thus conclude that the physical subspace associated to the classical BRST differential $s$ 
is not invariant under the application of the S matrix.

\vskip 0.4 truecm

As a final point of this section, we notice that
the cohomology of a nilpotent differential $\delta$
is known to be independent of the doublet $(z,w)$,
$\delta z = w \, , \delta w = 0$
whenever \cite{PS}
\begin{eqnarray}
[\delta, {\cal N}] = 0
\label{d5}
\end{eqnarray}
where ${\cal N}$ is the counting operator for the doublet $(z,w)$:
\begin{eqnarray}
{\cal N} = \int d^4x \, \left ( z {\delta \over \delta z}
+ w {\delta \over \delta w} \right ) \, .
\label{d6}
\end{eqnarray}
The analysis carried out in this section
shows that this result cannot be extended to more
general situations where eq.(\ref{d5}) is not fulfilled:
 for the extended linearized ST operator
${\cal S}'_0$  eq.(\ref{d5}) is not satisfied and
the cohomology of ${\cal S}'_0$ in the FP neutral sector
is actually $K$-dependent.

\section{Conclusions}

In this paper we have performed the quantization of chiral QED with one family of 
massless fermions.
We have introduced the Stueckelberg field in order to give mass to the
Abelian gauge field in a BRST invariant way and we have shown 
that an extended ST identity can be introduced
and fulfilled to all orders in perturbation theory by a suitable choice
of the local action-like counterterms, order by order in the loop-wise
expansion.

This ST identity incorporates the Adler-Bardeen anomaly and involves the introduction 
of an external source $K$ of dimension zero. By a purely algebraic argument
we have shown that the introduction of the source $K$ trivializes
the cohomology of the extended 
linearized classical ST operator ${\cal S}'_0$ 
in the FP charge $+1$ sector.

We have then discussed the physical content of the extended ST identity.
We have shown that the cohomology classes associated with ${\cal S}'_0$ are modified
with respect to the ones of ${\cal S}_0$. 
This provides a counterexample showing
that, if the counting operator for the doublet $(z,w)$ does not commute with the nilpotent differential
$\delta$ under which $(z,w)$ form a doublet, the cohomology of $\delta$ actually depends on $(z,w)$.

Hence the local physics generated by ${\cal S}'_0$ is modified with respect
to the one issued from ${\cal S}_0$. Since the latter is the same as the one generated from the
classical BRST differential, the physical states corresponding to $s$  do not survive quantization.
We have explicitly checked this result by showing that the subspace of the physical states
corresponding to $s$ is not left invariant under the application of the S matrix,
as a consequence of the extended ST identity satisfied by the quantum effective action $\G$.

\subsection*{Acknowledgments}
We acknowledge a partial financial support by MURST.

\end{document}